# Simple MaxEnt Models for Food Web Degree Distributions

Richard J. Williams

*Microsoft Research, 7 J J Thomson Ave, Cambridge CB3 0FB, UK*

E-mail: ricw@microsoft.com

**Abstract**

Degree distributions have been widely used to characterize biological networks including food webs, and play a vital role in recent models of food web structure. While food webs degree distributions have been suggested to follow various functional forms, to date there has been no mechanistic or statistical explanation for these forms. Here I introduce models for the degree distributions of food webs based on the principle of maximum entropy (MaxEnt) constrained by the number of species, number of links and the number of top or basal species. The MaxEnt predictions are compared to observed distributions in 51 food webs. The distributions of the number of consumers and resources in 23 (45%) and 35 (69%) of the food webs respectively are not significantly different at a 95% confidence level from the MaxEnt distribution. While the resource distributions of niche model webs are well-described by the MaxEnt model, the consumer distributions are more narrowly distributed than predicted by the MaxEnt model. These findings offer a new null model for the most probable degree distributions in food webs. Having an appropriate null hypothesis in place allows informative study of the deviations from it; for example these results suggest that there is relatively little pressure favoring generalist versus specialist consumption strategies but that there is more pressure driving the consumer distribution away from the MaxEnt form. Given the methodological idiosyncrasies of current food web data, further study of such deviations will need to consider both biological drivers and methodological bias.

**Introduction**

An enormous variety of strategies have evolved by which organisms capture the resources necessary for life, and by which organisms avoid being consumed as a resource. These strategies range from organisms that are specialized on a single resource species to ones that consume a wide range of resources at multiple trophic levels. Similarly, some organisms have evolved elaborate defensive strategies and are consumed by few species while others are vulnerable to a much wider range of consumers. The nature of the balance between specialization and generality in consumers, the range of vulnerability of resources, and the determination of the biological processes that drive these interrelationships are central problems in food web ecology (Dunne 2006).

Food web degree distributions, or the distribution of the fraction of nodes in a network with a particular number of links, provide a description of this balance. Degree distributions play a central role in the description and interpretation of the structure of complex networks (Strogatz 2001; Albert & Barabasi 2002) and have been widely used to characterize biological networks (Jordano *et al.* 2003; Barabasi & Oltvai 2004; May 2006) including food webs. They also play a vital role in recent models of food web structure (Stouffer *et al.* 2005). Despite their importance, to date there has been no mechanistic or statistical explanation for this vitally important aspect of food web structure.

A food web is a directed network of $S$ nodes connected by $L$ links, with links indicating the flow of biomass between nodes, which typically represent species or more coarsely resolved aggregations of species. Previous work on degree distributions in food webs has described their functional form. An early study of three food webs considered the undirected degree distribution, combining incoming and outgoing links, and suggested that degree distributions followed a power law and so are scale-free (Montoya & Sole 2002). This was disputed by a study of seven food webs, which considered the consumer and resource distributions separately and argued that both followed a single-scale functional form (Camacho *et al.* 2002). A study of 16 food webs found that the form of the undirected degree distributions varied with network connectance ($C = L/S^2$), with power law distributions at low values of connectance (Dunne *et al.* 2002). None of these studies provide any explanation as to why these distributions should occur.

In addition to their use in the description of complex networks, degree distributions play an important role in the performance of models of food web structure. Recently, it has been shown that the success of various structural models of complex food webs (the niche model (Williams & Martinez 2000) and its variants (Cattin *et al.* 2004; Stouffer *et al.* 2005; Stouffer *et al.* 2006; Williams & Martinez 2008)) depends in large part on the form of the resource distribution (Stouffer *et al.* 2005). While the other components of the niche model, ordering of species in a feeding hierarchy and constraining diets to contiguous niches, are grounded in well-established ecological ideas (Hutchinson 1959; Cohen 1978; Cohen *et al.* 1990), no justification was given for the resource distribution in the niche model, and this centrally important choice has simply been copied in more recent models.

Here I propose simple null models for the consumer, resource and undirected degree distributions of food webs which help fill this important gap in our understanding of food web structure. It has often been argued (Albert & Barabasi 2002; Montoya & Sole 2002; May 2006) that a random network (Erdős & Rényi 1959) where any link is equally probable is a suitable null model, with deviations in the degree distributions from the sharply-peaked binomial distribution of this model requiring explanation. This model assumes that all links occur with equal probability and therefore when considering the nodes in the network, it assumes that every node behaves identically; this assumption imposes biologically unlikely constraints on the degree distributions.

According to the principle of maximum entropy (MaxEnt) (Jaynes 1957), the probability distribution with the maximum information entropy is the least biased probability distribution which satisfies a set of information containing constraints. Here, I compare observed food web degree distributions to MaxEnt models constrained only by the numbers of species, top or basal species and links in the food webs. I also tested whether the degree distributions of niche model food webs (Williams & Martinez 2000) followed the MaxEnt models and whether deviations from the MaxEnt models were similar in the niche model and the empirical data.

**Materials and Methods.**

The consumer and resource distributions of the trophic species (Cohen *et al.* 1990) in 51 food webs were analyzed. The data are all the webs with 25 or more trophically distinct taxa (Cohen *et al.* 1990) from two recent studies (Stouffer *et al.* 2007; Thompson *et al.* 2007); details of the data are given in the supplementary information tables S1 and S2. These are among the largest and best resolved data available, and while still subject to the many criticisms that food web data have received (Cohen *et al.* 1993), the many robust patterns found in these methodologically heterogeneous data (Stouffer *et al.* 2007; Thompson *et al.* 2007; Williams & Martinez 2008) give confidence that these findings are not the result of consistent bias in the data.

Two resource distributions were considered, termed the "all-species resource distribution" and the "restricted resource distribution". The "all-species resource distribution" is defined as the distribution of the number of resources of each species, including the basal species, which consume no resources. This model is constrained only by knowledge of S and L. The "restricted resource distribution" is defined as the distribution of the number of resources of only the consumer species. As such, it includes prior knowledge of the number of basal species $B$ and does not attempt to predict the fraction of basal species. Similarly, two consumer distributions are considered, the "all-species consumer distribution" and the "restricted consumer distribution". The "all-species consumer distribution" is defined as the distribution of the number of consumers of each species, including the top species, which have no consumers. This model is constrained only by knowledge of S and L. The "restricted consumer distribution" is defined as the distribution of the number of consumers of the resource species, includes prior knowledge of the number of top species $T$ and does not attempt to predict the fraction of top species.

In the "all species" distributions, the number of consumers or resources of each species can range from 0 to S and the mean number of links per species is L/S. In the "restricted" resource distribution, the number of links to each consumer can potentially range from *1* to *S* and the mean number of links to each consumer is $L/(S - B)$. In the "restricted" consumer distribution, the number of links from each resource can potentially range from *1* to *S* and the mean number of links from each resource is $L/(S - T)$. For a discrete distribution on the set of values $\{x_1,\ldots,x_n\}$ with mean $\mu$, the MaxEnt distribution is (Cover & Thomas 2006) $p_i = P(X = x_i) = Ce^{\lambda x_i}$ for $i = 1,\ldots,n$. The constants *C* and $\lambda$ are determined by the requirements that the probabilities sum to 1 and have mean $\mu$: $\sum_i p_i = 1$ and $\sum_i x_i p_i = \mu$.

Finally, I developed a simple model of the undirected (sum of consumer and resource link) distributions by assuming that the number of consumers and resources of each node are independent. Top species have no consumers, so for *T* species the number of links is drawn from the MaxEnt resource distribution. Similarly, for *B* species the number of links is drawn from the MaxEnt consumer distribution. For the remaining $S - B - T$ intermediate species, the number of links is the sum of numbers drawn from the consumer and resource distributions.

The consumer, resource and undirected distributions of the 51 empirical food webs were compared to the maximum entropy distributions derived using the empirical values of *S*, *L*, *B* and *T*. Two tests of the fit of the MaxEnt models to the empirical data were used. In the first, the likelihood ratio (*G*) statistic (Sokal & Rohlf 1995) is used to compare an observed distribution to some expected (model) distribution. *G* is defined as $G = 2\sum_i O_i \ln(O_i / E_i)$ where $O_i$ is the observed frequency, $E_i$ the expected frequency and *i* indexes through all values in the discrete distribution with non-zero expected value. A randomization procedure is used; for each of the 10,000 trials a sample is drawn from the maximum entropy distribution and its *G* value is compared to the *G* value of the empirical distribution, where in both cases the expected distribution is the maximum entropy distribution. The goodness of fit $f_G$, is measured by the fraction of trials in which the *G* value of the empirical distribution is greater than the *G* value of the distribution drawn from the maximum entropy distribution. The empirical distribution is considered to be significantly different from the maximum entropy distribution if $f_G > 0.95$.

The goodness of fit $f_G$ does not differentiate between webs with overly broad or narrow degree distributions, a range of variation found in an earlier study of food web degree distributions (Dunne *et al.* 2002). To measure whether the empirical webs were more broadly or narrowly distributed than the model distributions, I measured the relative width of a distribution $W = log(\sigma_O/\sigma_M)$ where $\sigma_O$ is the standard deviation of the observed distribution and $\sigma_M$ is the standard deviation of the model distribution. For each empirical web, the distribution of *W* for 10,000 webs drawn from the model distribution was computed. The quantity $W_{95}$ is defined as the deviation of the empirical value of *W* from the model median normalized by the width of the upper or lower half of the central interval of the model distribution of *W* at the 95% significance level. This gives the normalized difference in standard deviations of the empirical distribution relative to the median standard deviation of a set of samples drawn from the model distribution and so measures the relative width of the empirical distribution. Webs with $W_{95} < -1$ have distributions that are significantly narrower than the model distributions; $W_{95} > 1$ occurs for distributions significantly broader than the model distributions.

**Results**

Table 1 summarizes the fit of the MaxEnt models to the empirical data using the two tests of model fit. An empirical distribution is significantly different from the model distribution at the 95% confidence level if $f_G > 0.95$. Webs with $W_{95} < -1$ or $W_{95} > 1$ are significantly narrower or broader than the model distributions respectively. The "all-species" models perform consistently worse than the models which are restricted to exclude nodes with zero links. The differences are much larger for resource distributions than for consumer distributions. This suggests that the number of basal species is particularly different from the number predicted by the all-species MaxEnt model, and a biological or methodological basis for their abundance should be sought. All subsequent results will be for the better performing "restricted" models which incorporate prior knowledge of the number of top or basal species.

In the most conservative evaluation, the restricted consumer and resource distributions are not significantly different from the model distribution at a 95% confidence level if both $f_G < 0.95$ and $-1 < W_{95} < 1$. These conditions are satisfied for 23 and 35 (45% and 69%) of the webs respectively. Thus there is some asymmetry between the fit of the consumer and resource distributions to their respective MaxEnt distributions ($p = 0.027$, Fisher's Exact Test). Many of the poorly fit degree distributions are only marginally significantly different from the MaxEnt model. Of the distributions with with $f_G > 0.95$, 12 of 23 consumer distributions and 4 of 9 resource distributions' $f_G$ fall between 0.95 and 0.99. The table also shows that the random model (Erdős & Rényi 1959) is a very poor predictor of the empirical degree distributions.

Figure 1 shows three examples of empirical degree distributions and the range of degree distributions found in 10,000 samples drawn from the corresponding MaxEnt model; data sets were selected to illustrate the range of variation found in the empirical degree distributions. (The full set of the empirical and model distributions are available as figure S1 in the supplementary information). The MaxEnt model fits the resource distribution of the StMartin web (figure 1a) quite closely but is a poor fit to the two other degree distributions shown. The consumer distribution of the Powder web (figure 1b) is more broadly distributed than the MaxEnt distributions, having more species with few consumers (low vulnerability) and more species with large numbers of consumers (high vulnerability) than predicted by the MaxEnt model. In contrast, the consumer distribution of the Reef web (figure 1c) is relatively narrowly distributed, with fewer either highly vulnerable or invulnerable species than predicted by the MaxEnt model. Note also that for the relatively high connectance Reef web (fig 1c), the MaxEnt distribution is truncated (curves downward) at large numbers of links compared to the near-exponential behavior (linear in the log-linear plots) of the lower connectance degree distributions (figs 1a,b).

In 16 webs both distributions are well fit by the MaxEnt models, in 19 webs only the resource distribution is well fit, in 7 webs only the consumer distribution is well fit and in 9 webs neither distribution is well fit. Fisher's Exact Test suggests that the two degree distributions are independent ($p = 1$). Given this result, I created a model for the undirected degree distribution by assuming that each node's incoming and outgoing links were drawn from independent MaxEnt models. Using the conditions that both $f_G < 0.95$ and $-1 < W_{95} < 1$, the undirected degree distributions were well-fit in 28 (57%) of the empirical webs. This result is intermediate between the results for the consumer and resource distributions taken separately and further reinforces the idea that the consumer and resource distributions can be treated as independent.

The fit of each web is shown in figure 2, which plots the relative width $W_{95}$ of the consumer and resource distributions of the empirical webs against goodness of fit $f_G$. Webs with poorly fit consumer and resource distributions ($f_G > 0.95$) have a wide range of relative widths, but are generally more broadly spread (positive $W_{95}$) than the MaxEnt model. Webs in figure 2 are broken into two groups, the stream webs collected by Thompson and his collaborators (Thompson & Townsend 2003, 2004) and all the other webs. There is a well-defined cluster of Thompson's stream webs whose consumer degree distributions are relatively broad and poorly fit by the MaxEnt model. The resource distributions of these webs are better fit by the MaxEnt model and while mostly not significantly different in width from the model webs, they do stand out as having relatively narrow distributions, indicated by their consistently negative values of $W_{95}$.

Goodness of fit ($f_G$) and relative width ($W_{95}$) of the resource distribution does not depend on network size ($S$) or mean connectivity ($L/S$). There is a weak, marginally significant relationship between the consumer distribution's $f_G$ and $S$ and a more strongly significant decrease in consumer $W_{95}$ with $L/S$, as shown in figure 3 (Details are given in the supplementary information table S3). This figure also shows the relatively broad consumer distributions of the Thompson stream webs. At higher $L/S$, the consumer distributions of the empirical webs tend to be narrower than the MaxEnt model distributions, with a more rapid drop-off at higher link values than predicted by the model. The truncation of the consumer distributions with increasing $L/S$ is more extreme than the truncation of the MaxEnt distribution at higher $L/S$ noted earlier.

These results, along with the strong correlation between $L/S$ and $C$ in these data, suggest that the truncation of the consumer distribution at higher $L/S$ drives the truncation of the undirected degree distribution at high $C$ noted in an earlier study (Dunne *et al.* 2002).

The same tests of goodness of fit were used to test whether the degree distributions of the niche model (Williams & Martinez 2000) followed these MaxEnt models and whether deviations from the MaxEnt models were similar in the niche model and the empirical data. Figure 4a,b shows the fraction of 1000 niche model food webs whose consumer distributions were significantly different from the MaxEnt model at the 95% confidence level using both the goodness of fit ($f_G$) and relative width ($W_{95}$) tests and the mean $f_G$ and $W_{95}$ of the niche model webs. Figure 4c,d shows the same information for the niche model resource distributions. These figures show that while the niche model resource distributions were always fairly close to the MaxEnt model, the niche model consumer distributions were consistently much more narrowly distributed than MaxEnt model. As $L/S$ increases, the empirical webs' consumer distributions tend to become more narrowly constrained than in the MaxEnt model (figure 3) but this trend is far stronger in the niche model (figure 4b). Finally, while the niche model resource distribution is reasonably well-fit by the MaxEnt model, the fit consistently worse (higher $f_G$) and the distribution is consistently broader (higher $W_{95}$) as the network increases in size (S) (figure 5). No such scale dependence is apparent in the empirical data.

**Discussion**

Two important pieces of information characterize the distribution of links in a food web, the total number of links in the system and hence the mean number of links per species, and the distribution of those links among the species in the food web, here characterized by the various degree distributions. This work does not attempt to explain the mean diet breadth (Beckerman *et al.* 2006; Petchey *et al.* 2008), or number of links per species, but instead addresses the drivers of the distribution about this mean. The relatively close agreement between the degree distributions of the 51 empirical food webs and the MaxEnt models shows that in many food webs, one does not need to consider detailed ecological processes to be able to predict the consumer, resource or undirected degree distributions.

When significant deviations from the MaxEnt distributions occur, other constraints are at work in determining the form of the degree distributions. When this occurs, closer examination of the ecological processes or observational techniques must be carried out to determine what processes are forcing the consumer or resource distribution away from the MaxEnt form.

One example of such a deviation is that the consumer distributions of the empirical webs are consistently narrower than the MaxEnt model empirical food webs, especially at high $L/S$ (figure 3). These distributions have a shorter tail (figure 1c), and so there are fewer taxa with large numbers of consumers than predicted by the MaxEnt model. A possible explanation for this is that the top-down pressure on taxa with large numbers of consumers increases their risk of extinction, and so empirical networks have fewer highly vulnerable taxa than predicted by the simple MaxEnt null model.

The results presented here also show that the Thompson stream webs have consistently broader consumer distributions than the MaxEnt model distributions. These webs comprise the vast majority of the stream webs analyzed, and some features of the ecology of stream habitats might cause this consistent difference in stream food web consumer distributions. It is also possible that the data gathering techniques used produced food webs that are consistently different from food webs generated using other techniques, as suggested in an earlier study (Stouffer *et al.* 2007). These webs stand out methodologically, being based on gut content analysis of a relatively small number of individuals of each species, leading to an acknowledged likely undersampling of links (Thompson & Townsend 2003). If rare links tend to be to relatively invulnerable species, increased sampling could make the consumer distributions less broadly distributed by reducing the number of species with very low vulnerability.

The comparison of the degree distributions of niche model food webs with the MaxEnt models show that there are a number of consistent differences between the degree distributions of the empirical webs and the niche model which need to be addressed by future structural models of food webs. Resource distributions are more narrowly distributed than predicted by the MaxEnt model and there is pronounced scale-dependence in the fit of the consumer distributions, with larger niche model webs having more broadly distributed consumer distributions.

Given the methodological variability of the data sets, not only between the Thompson data and the other webs but also across the other webs (Dunne *et al.* 2004; Stouffer *et al.* 2007), the degree distributions of complex food webs are remarkably well described by the simple MaxEnt model presented here. The many questions surrounding data quality mean that it is currently difficult to assess whether deviations from the MaxEnt model are a result of ecological processes or biases in the data.

**Acknowledgements**

Thanks to Jen Dunne and Ross Thompson for generously sharing their data sets, to Jen Dunne, Drew Purves and David Stouffer for helpful discussions and comments on earlier versions of this paper.

**References**

Albert R. & Barabasi A.L. (2002). Statistical mechanics of complex networks. *Rev. Modern Phys.*, 74, 47-97.

Barabasi A.L. & Oltvai Z.N. (2004). Network Biology: Understanding the Cell's Functional Organization. *Nature Reviews Genetics*, 5, 101-113.

Beckerman A.P., Petchey O.L. & Warren P.H. (2006). Foraging biology predicts food web complexity. *Proceedings of the National Academy of Science, USA*, 103, 13745-13749.

Camacho J., Guimera R. & Amaral L.A.N. (2002). Robust patterns in food web structure. *Phys. Rev. Lett.*, 88.

Cattin M.-F., Bersier L.-F., Banasek-Richter C., Baltensperger R. & Gabriel J.-P. (2004). Phylogenetic constraints and adaptation explain food-web structure. *Nature*, 427, 835-839.

Cohen J.E. (1978). *Food Webs and Niche Space*. Princeton University Press, Princeton, N.J.

Cohen J.E., Beaver R.A., Cousins S.H., DeAngelis D.L., Goldwasser L., Heong K.L., Holt R.D., Kohn A.J., Lawton J.H., Martinez N., O'Malley R., Page L.M., Patten B.C., Pimm S.L., Polis G.A., Rejm nek M., Schoener T.W., Schoely K., Sprules W.G., Teal J.M., Ulanowicz R.E., Warren P.H., Wilbur H.M. & Yodzis P. (1993). Improving food webs. *Ecology*, 74, 252-258.

Cohen J.E., Briand F. & Newman C.M. (1990). *Community food webs: data and theory*. Springer, Berlin.

Cover T.M. & Thomas J.A. (2006). *Elements of Information Theory (2nd Edition)*. Wiley-Interscience.

Dunne J.A. (2006). The network structure of food webs. In: *Ecological Networks: Linking Structure to Dynamics in Food Webs* (eds. Pascual M & Dunne JA). Oxford University Press New York.

Dunne J.A., Williams R.J. & Martinez N.D. (2002). Food-web structure and network theory: The role of connectance and size. *Proc. Natl Acad. Sci. USA*, 99, 12917-12922.

Dunne J.A., Williams R.J. & Martinez N.D. (2004). Network structure and robustness of marine food webs. *Mar Ecol Prog Ser*, 273, 291-302.

Erdős P. & Rényi A. (1959). On random graphs I. *Publicationes Mathematicae Debrecen*, 6, 290-297.

Hutchinson G.E. (1959). Homage to Santa Rosalia or Why are there so many kinds of animals? *Am. Nat.*, 93, 145-159.

Jaynes E.T. (1957). Information theory and statistical mechanics. *Physical Review*, 106, 620-630.

Jordano P., Bascompte J. & Olesen J.M. (2003). Invariant properties in coevolutionary networks of plant-animal interactions. *Ecol Lett*, 6, 69-81.

May R.M. (2006). Network structure and the biology of populations. *Trends Ecol. Evol.*, 21, 394-399.

Montoya J.M. & Sole R.V. (2002). Small world patterns in food webs. *J. Theor. Biol.*, 214, 405-412.

Petchey O.L., Beckerman A.P., Riede J.O. & Warren P.H. (2008). Size, foraging and food web structure. *Proceedings of the National Academy of Science, USA*, 105, 4191-4196.

Sokal R.R. & Rohlf F.J. (1995). *Biometry*. Freeman, New York.

Stouffer D.B., Camacho J. & Amaral L.A.N. (2006). A robust measure of food web intervality. *Proceedings of the National Academy of Science, USA*, 103, 19015-19020.

Stouffer D.B., Camacho J., Guimera R., Ng C.A. & Amaral L.A.N. (2005). Quantitative patterns in the structure of model and empirical food webs. *Ecology*, 86, 1301-1311.

Stouffer D.B., Camacho J., Jiang W. & Amaral L.A.N. (2007). Evidence for the existence of a robust pattern of prey selection in food webs. *Proc R Soc Lond B*, 274, 1931-1940.

Strogatz S.H. (2001). Exploring complex networks. *Nature*, 410, 268-276.

Thompson R.M., Hemberg M., Starzomski B.M. & Shurin J.B. (2007). Trophic levels and trophic tangles: the prevalence of omnivory in real food webs. *Ecology*, 88, 612-617.

Thompson R.M. & Townsend C.R. (2003). Impacts on stream food webs of native and exotic forest: An intercontinental comparison. *Ecology*, 84, 145-161.

Thompson R.M. & Townsend C.R. (2004). Landuse influences on New Zealand stream communities – effects on species composition, functional organization and food-web structure. *New Zealand Journal Marine and Freshwater Research*, 38, 595–608.

Williams R.J. & Martinez N.D. (2000). Simple rules yield complex food webs. *Nature*, 404, 180-183.

Williams R.J. & Martinez N.D. (2008). Success and its limits among structural models of complex food webs. *J. Anim. Ecol.*

**Table 1.** Number and (Fraction) of 51 food webs which are not significantly different from the MaxEnt and binomial (random model) distribution based on various criteria.

| Criteria | All-Species Consumer Distr | All-Species Resource Distr |
|---|---|---|
| $f_G < 0.95$ | 25 (0.49) | 21 (0.41) |
| $W_{95} > -1$ and $W_{95} < 1$ | 28 (0.55) | 41 (0.80) |
| $W_{95} > -1$, $W_{95} < 1$ and $f_G < 0.95$ | 21 (0.41) | 20 (0.39) |
| $f_G < 0.99$ | 36 (0.71) | 28 (0.55) |
| | | |
| Binomial $f_G < 0.99$ | 1 (0.02) | 4 (0.078) |
| | | |
| | Restricted Consumer Distr | Restricted Resource Distr |
| $f_G < 0.95$ | 28 (0.55) | 42 (0.82) |
| $W_{95} > -1$ and $W_{95} < 1$ | 31 (0.61) | 40 (0.78) |
| $W_{95} > -1$, $W_{95} < 1$ and $f_G < 0.95$ | 23 (0.45) | 35 (0.69) |
| $f_G < 0.99$ | 39 (0.76) | 47 (0.92) |

**Figure 1. Empirical and MaxEnt Cumulative Degree Distributions of Three Food Webs.** Linear-log plots of the cumulative distribution (fraction of taxa with more than $k$ consumers or resources). On the x-axis, $k$ is scaled by the mean number of links, $z = L/S$. Red curve is the empirical distribution, black curve the mean and grey curves the upper and lower limits of the central 95% of 10,000 distributions drawn from the MaxEnt model. Inset in each panel shows the food web's name, distribution type, connectance C, likelihood ratio goodness of fit $f_G$ and the relative width $W_{95}$.

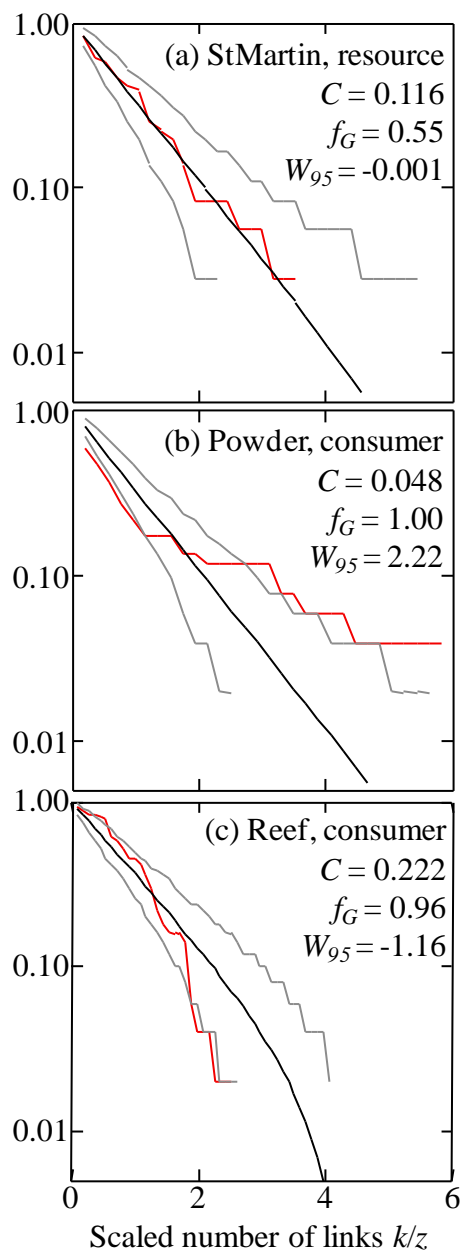

**Figure 2. Goodness of Fit of Empirical Food Webs to MaxEnt Model.** Likelihood ratio goodness of fit $f_G$ plotted against the relative width of the empirical distribution $W_{95}$ for (a) consumer and (b) resource distributions. Thompson and collaborator's stream webs are indicated by ×. Points that fall above the upper and below the lower dotted horizontal lines have distribution widths significantly broader and narrower at the 95% confidence level than the MaxEnt distribution respectively. Points to the right of the vertical dotted line have significantly low likelihood at the 95% confidence level of being drawn from the MaxEnt distribution.

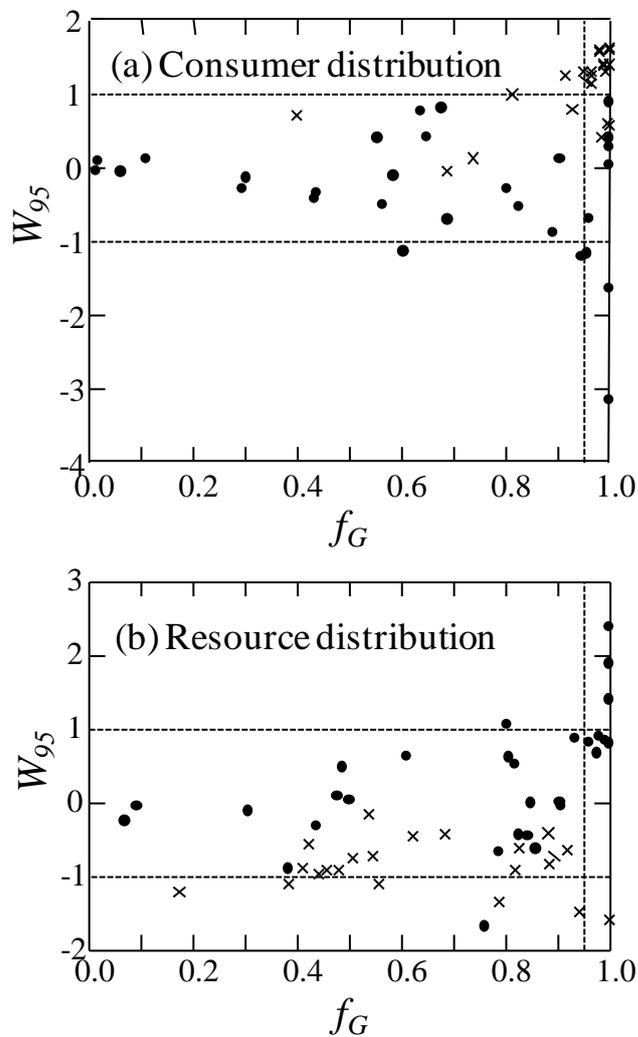

**Figure 3. Dependence of Consumer Distribution Width on Connectance.** Relative width of the empirical consumer distribution $W_{95}$ plotted against mean connectivity L/S. Thompson and collaborator's stream webs are indicated by ×. Points that fall above the upper and below the lower dotted horizontal lines have distribution widths significantly broader and narrower at the 95% confidence level than the MaxEnt distribution respectively.

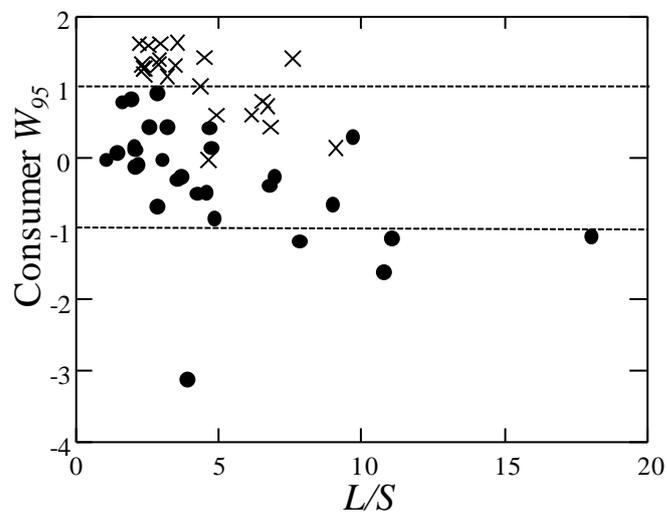

**Figure 4. Goodness of Fit and Relative Width of Niche Model Food Webs.**
Fraction of 1000 niche model food webs that are outside the 95% confidence interval for goodness of fit ($f_G$) and relative width ($W_{95}$) for the (a) consumer distribution versus $L/S$, (c) resource distribution versus $L/S$; and the mean value of $f_G$ and $W_{95}$ for 1000 niche model food webs for the (b) consumer distribution versus $L/S$, (d) resource distribution versus $L/S$.

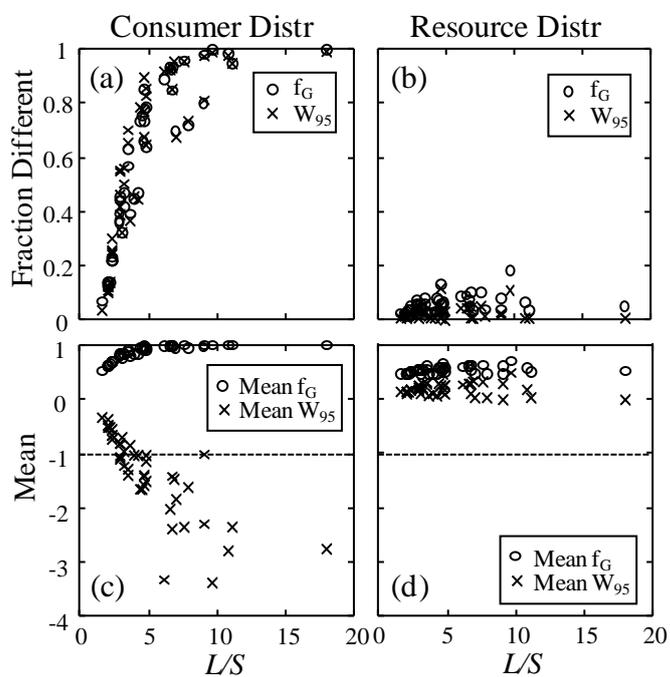

**Figure 5. Scale Dependence of the Goodness of Fit and Relative Width of the Resource Distribution of Niche Model Food Webs.** Mean value of fG and W95 of the resource distribution versus S. Each point is the mean across a set of 1000 niche model food webs with the same S and C as an empirical food web.

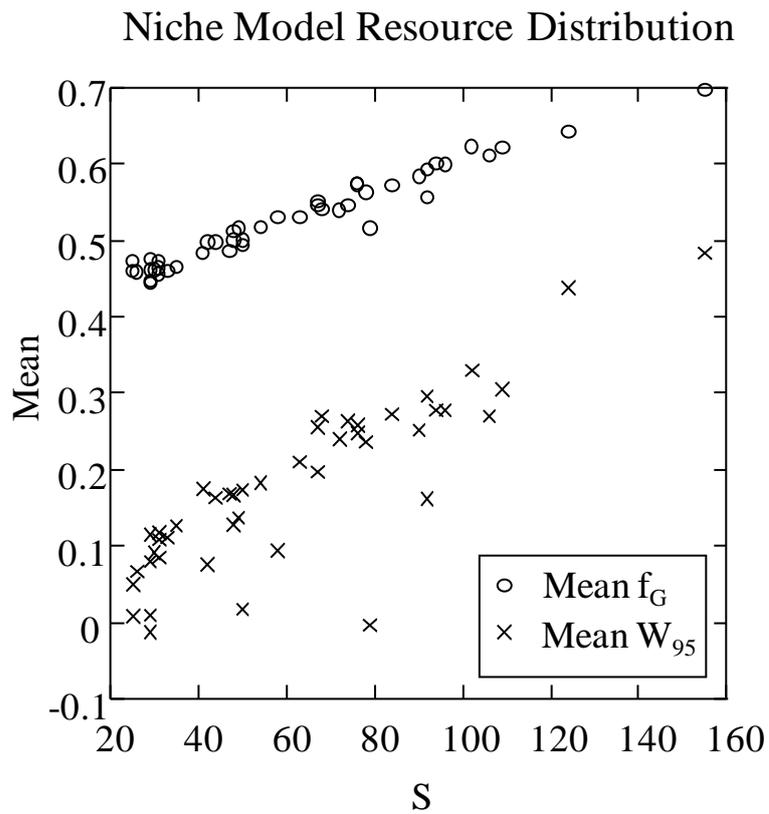

22**Supplementary Information for "Simple MaxEnt Models for Food Web Degree Distributions"**

**Table S1. Data Sets**

| Web Name | Source | S | L/S | C |
|---|---|---|---|---|
| AireStream | 1 | 49 | 2.898 | 0.059 |
| CrocodileCreek | 1 | 29 | 1.655 | 0.057 |
| DeepCreek | 1 | 26 | 3.731 | 0.143 |
| LakeNyasa1 | 1 | 31 | 3.065 | 0.099 |
| LakeNyasa2 | 1 | 33 | 2.121 | 0.064 |
| OakGall | 1 | 54 | 3.222 | 0.06 |
| Saltmeadow | 1 | 32 | 1.094 | 0.034 |
| ShortgrassPrairie | 1 | 106 | 3.575 | 0.034 |
| SonoranDesert | 1 | 48 | 2.875 | 0.06 |
| TreleaseWoods | 1 | 29 | 2.103 | 0.073 |
| MontereyBay | 1 | 35 | 2.086 | 0.06 |
| Benguela | 2 | 29 | 7 | 0.241 |
| Broom | 3 | 85 | 2.624 | 0.031 |
| ElVerde | 4 | 155 | 9.735 | 0.063 |
| StMartin | 5 | 42 | 4.881 | 0.116 |
| BridgeBrookLake | 6 | 25 | 4.28 | 0.171 |
| UKGrassland | 7 | 73 | 1.507 | 0.021 |
| LittleRockLake | 8 | 92 | 10.837 | 0.118 |
| Coachella | 9 | 29 | 9.034 | 0.312 |

| Name | | | | |
|---|---|---|---|---|
| Reef | 10 | 50 | 11.12 | 0.222 |
| Shelf | 11 | 79 | 18.076 | 0.229 |
| StMarks | 12 | 48 | 4.604 | 0.096 |
| ChesapeakeBay | 13 | 31 | 2.226 | 0.072 |
| SkipwithPond | 14 | 25 | 7.88 | 0.315 |
| Ythan | 15 | 78 | 4.795 | 0.061 |
| Ythan96 | 16 | 124 | 4.702 | 0.038 |
| Blackrock | 17 | 84 | 4.381 | 0.052 |
| Broad | 17 | 90 | 6.167 | 0.069 |
| Canton | 17 | 102 | 6.833 | 0.067 |
| German | 17 | 76 | 4.513 | 0.059 |
| Healy | 17 | 94 | 6.713 | 0.071 |
| KyeBurn | 17 | 96 | 6.531 | 0.068 |
| LittleKyeBurn | 17 | 76 | 4.908 | 0.065 |
| Dempters | 17 | 106 | 9.104 | 0.086 |
| Stony | 17 | 109 | 7.606 | 0.07 |
| Sutton | 17 | 67 | 4.657 | 0.07 |
| AkatoreA | 17 | 72 | 2.917 | 0.041 |
| AkatoreB | 17 | 44 | 2.364 | 0.054 |
| Berwick | 17 | 68 | 3.191 | 0.047 |
| CompanyBay | 18 | 58 | 6.81 | 0.117 |
| Coweeta1 | 19 | 47 | 2.319 | 0.049 |
| Coweeta17 | 19 | 58 | 2.241 | 0.039 |
| DuffinCreek | 20 | 30 | 3.933 | 0.131 |
| Lerderderg | 21 | 31 | 1.968 | 0.063 |
| Martins | 19 | 92 | 3.5 | 0.038 |
| Mimihau | 17 | 63 | 2.889 | 0.046 |
| Narrowdale | 17 | 50 | 2.4 | 0.048 |
| NorthCol | 17 | 67 | 2.97 | 0.044 |
| Powder | 17 | 74 | 3.554 | 0.048 |
| Troy | 19 | 63 | 2.54 | 0.04 |
| Wisp | 17 | 41 | 2.39 | 0.058 |

**Table S2. Data Set Sources**

| | |
|---|---|
| 1 | Cohen, J. E. 1989. Ecologists' Co-operative Web Bank ( ECOWeB), Version 1.0 (machine-readable database). Rockefeller University, New York, New York, USA. |
| 2 | Yodzis, P. 1998. Local trophodynamics and the interaction of marine mammals and fisheries in the Benguela ecosystem. J. Anim. Ecol. 67, 635. |
| 3 | Hawkins, B. A., N. D. Martinez, and F. Gilbert. 1997. Source food webs as estimators of community web structure. International Journal of Ecology 18:575–586. |
| 4 | Waide, R. B., and W. B. Reagan. Editors. 1996. The food web of a tropical rainforest. University of Chicago Press, Chicago, Illinois, USA |
| 5 | Goldwasser, L., and J. A. Roughgarden. 1993. Construction of a large Caribbean food web. Ecology 74:1216–1233. |
| 6 | Havens, K. 1992. Scale and structure in natural food webs. Science 257:1107–1109 |
| 7 | Memmott, J., N. D. Martinez, and J. E. Cohen. 2000. Predators, parasites and pathogens: species richness, trophic generality, and body sizes in a natural food web. Journal of Animal Ecology 69:1–15. |
| 8 | Martinez, N. D. 1991. Artifacts or attributes? Effects of resolution on the Little Rock Lake food web. Ecological Monographs 61:367–392 |
| 9 | Polis, G. A. 1991 Complex trophic interactions in deserts: an empirical critique of food-web theory. American Naturalist 138:123–155 |
| 10 | Opitz, S. 1996. Trophic interactions in Caribbean coral reefs. ICLARM [International Center for Living Aquatic Resources Management] Technical Reports 43, 341. |
| 11 | Link, J. 2002. Does food web theory work for marine ecosystems? Mar Ecol Prog Ser 230, 1. |
| 12 | Christian, R. R., and J. J. Luczkovich. 1999. Organizing and understanding a winter's seagrass foodweb network through effective trophic levels. Ecological Modelling 117:99–124. |
| 13 | Baird, D., and R. E. Ulanowicz. 1989. The seasonal dynamics of the Chesapeake Bay ecosystem. Ecological Monographs 59:329–364. |
| 14 | Warren, P.H. (1989). Spatial and temporal variation in the structure of a freshwater food web. Oikos, 55, 299–311. |
| 15 | Hall, S. J., and D. Raffaelli. 1991. Food-web patterns: lessons from a species-rich web. Journal of Animal Ecology 60:823–842. |
| 16 | Huxham, M., S. Beany, and D. Raffaelli. 1996. Do parasites reduce the chances of triangulation in a real food web? Oikos 76, 284. |
| 17 | Thompson, R. M., and C. R. Townsend. 2004 Landuse influences on New Zealand stream communities – effects on species composition, functional organization and food-web structure. New Zealand Journal Marine and Freshwater Research 38:595–608. |
| 18 | Thompson, R. M., K. Mouristen, and R. Poulin. 2005. Importance of parasites and their life cycle characteristics in determining the structure of a large marine food web. Journal of Animal Ecology 74:77–85. |
| 19 | Thompson, R. M., and C. R. Townsend. 2003. Impacts on stream food webs of native and exotic forest: an intercontinental comparison. Ecology 84:145–161. |
| 20 | Tavares-Cromar, A. F., and D. D. Williams. 1996 The importance of temporal resolution in food web analysis: Evidence from a detritus-based stream. Ecological Monographs 66:91–113. |

| 21 | Closs, G. P., and P. S. Lake. 1994 Spatial and temporal variation in the structure of an intermittent stream food web. Ecological Monographs 64:1–21. |

**Table S3. Linear regression of goodness of fit $f_G$ and relative width $W_{95}$ versus S and L/S for both consumer and resource networks.**

| Variable | Independent Var | $R^2$ | p | Slope |
|---|---|---|---|---|
| Consumer $f_G$ | S | 0.084 | 0.039 | |
| | L/S | 0.030 | 0.223 | |
| Resource $f_G$ | S | 0 | 0.953 | |
| | L/S | 0.015 | 0.384 | |
| Consumer $W_{95}$ | S | 0.065 | 0.071 | |
| | L/S | 0.179 | 0.002 | -0.132 |
| Resource $W_{95}$ | S | 0.025 | 0.264 | |
| | L/S | 0 | 0.961 | |

**Figure S1. Cumulative predator and prey distributions of the 51 empirical food webs.** Linear-log plots of the cumulative distribution (fraction of species (nodes) with more than $k$ predators or prey. On the x-axis, the number of links is scaled by the mean number of links, $z = L/S$. Red curce is the empirical degree distribution, black curve the mean and grey curves the upper and lower limits of the central 95% of 10,000 distributions drawn from the Maxent model. Inset in each panel shows the food web's connectance C, likelihood ratio goodness of fit $f_G$ and fit measured by the relative width of the distribution $W_{95}$.

# Cumulative Consumer Distributions

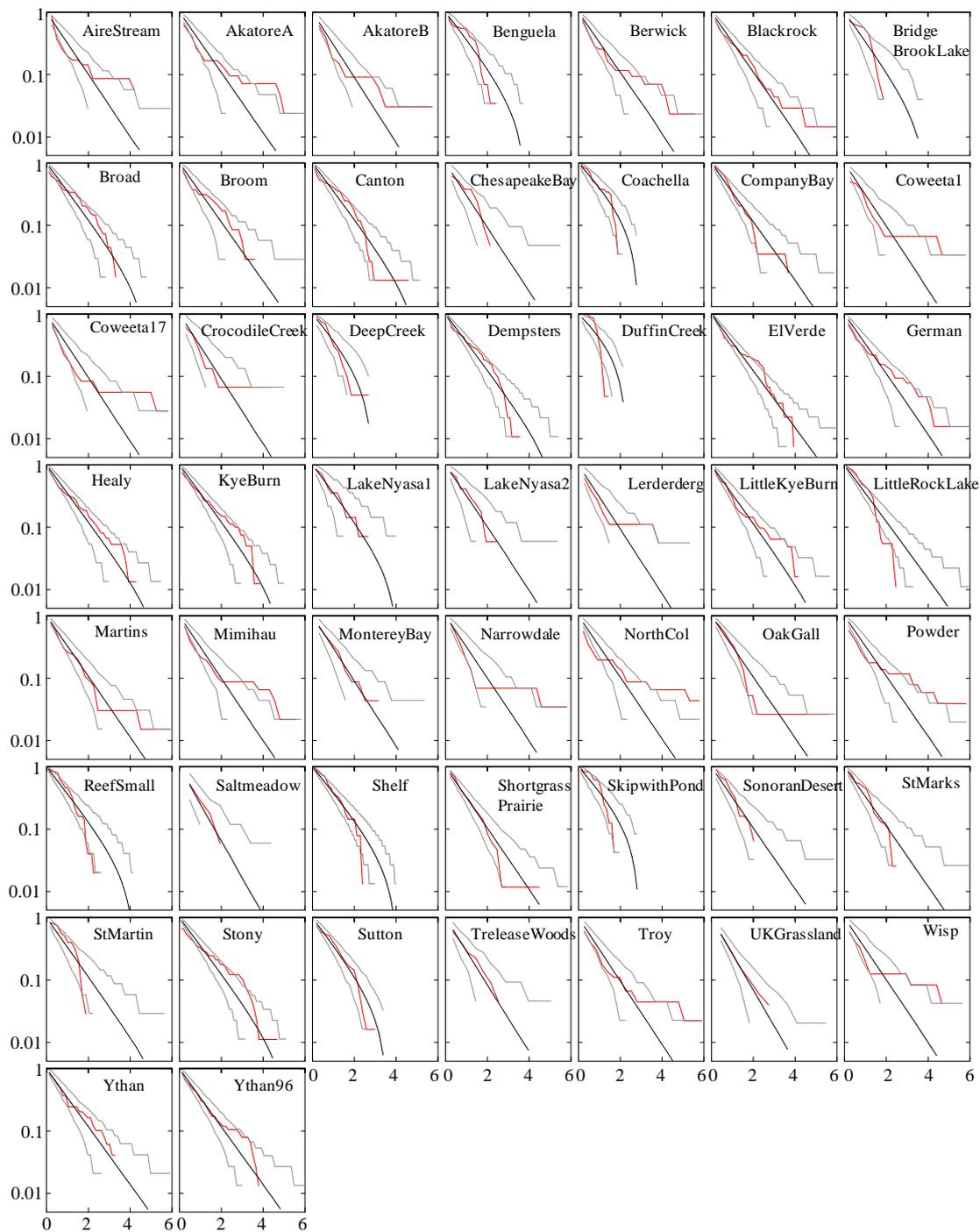
28
28

# Cumulative Resource Distributions

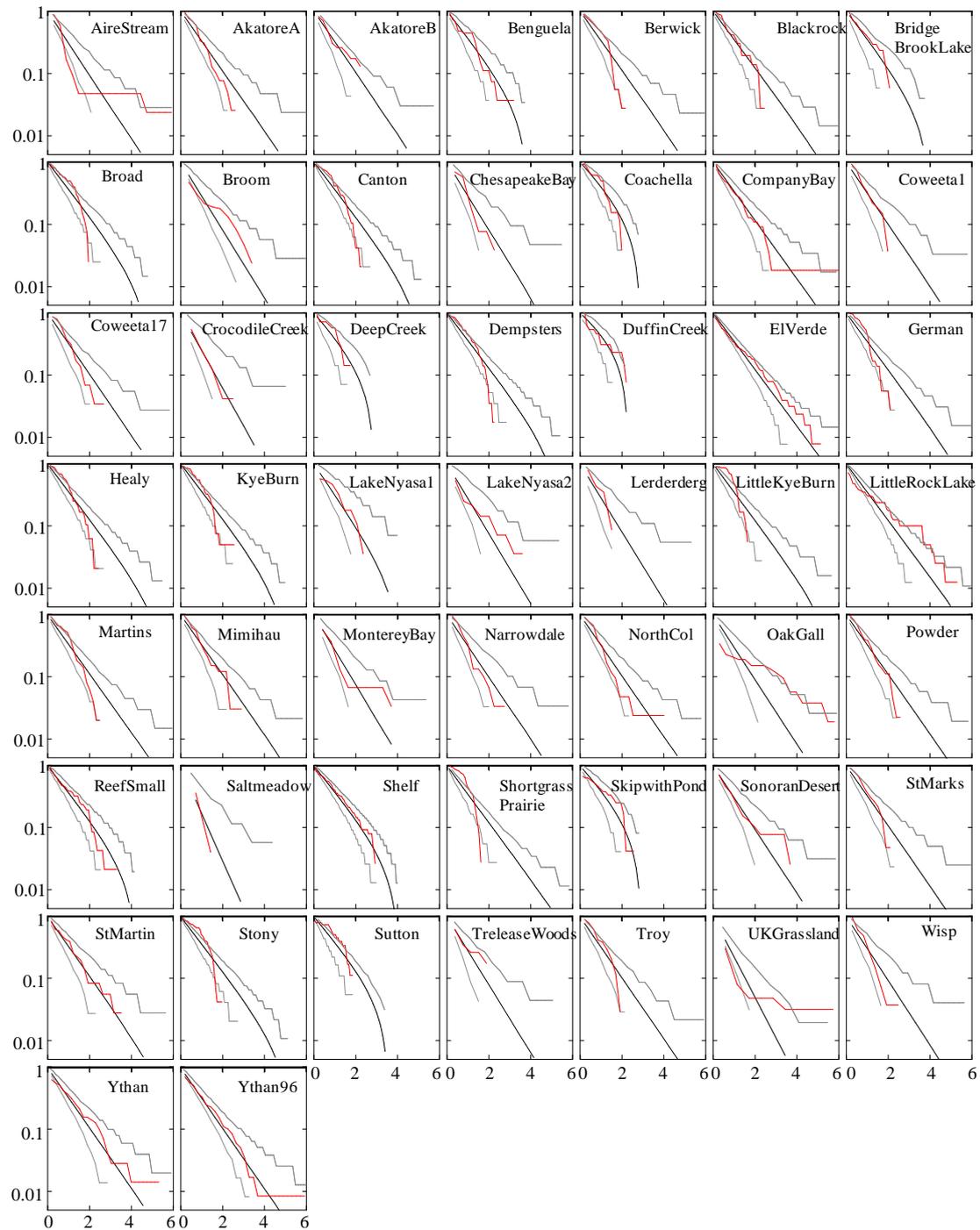